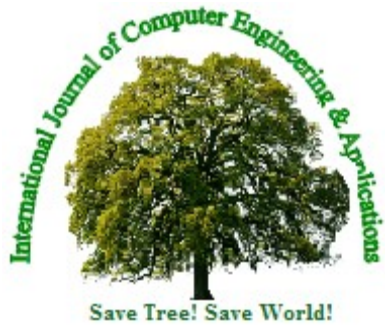

# SURVEY ON VARIANTS OF DISTRIBUTED ENERGY EFFICIENT CLUSTERING PROTOCOLS IN HETEROGENEOUS WIRELESS SENSOR NETWORK

**Manpreet Kaur [1], Abhilasha [2]**

*Department of Computer Science and Engineering*
*Punjab Technical University Gaini Zail Singh Campus*
*Bathinda, Punjab*

**ABSTRACT:**

*Wireless sensor networks are composed of low cost and extremely power constrained sensor nodes which are scattered over a region forming self organized networks, making energy consumption a crucial design issue. Thus, finite network lifetime is widely regarded as a fundamental performance bottleneck. These networks are used for various applications such as field monitoring, home automation, medical data collection or surveillance. Research has shown that clustering sensor nodes is an efficient method to manage energy consumption for prolonging the network lifetime. Presence of heterogeneity enhances the lifetime and reliability in network. In this paper, we present the distributed and energy efficient clustering protocols which follow the thoughts of Distributed Energy Efficient Clustering (DEEC) protocol. Objective of our work is to analyze that how these extended routing protocols work in order to optimize network lifetime and how routing protocols are improved. We emphasizes on issues experienced by various protocols and how these issues are tackled by other enhanced protocols. This provides a survey of work done on distributed energy efficient protocols.*

## [1] INTRODUCTION

Since the start of the third Millennium, potential use of wireless sensor network (WSN) has increased in various fields due to their applicability in various fields like security, environments, health and agriculture etc. Network is usually consists of one special node spatially distributed with unlimited battery life called sink or base station (BS). Sink gathers the data from low-power sensor nodes which are deployed in sensor field, process it and reports to user. Sensor field is observation area where multiple sensor nodes communicate by wireless radio signals. Functions performed by sensor nodes it very useful for Acoustic detection, Seismic Detection, Military surveillance, Inventory tracking, Medical monitoring, Smart spaces, Process Monitoring. Sensor nodes are the building blocks of network. Sensor nodes are usually built up of memory elements and a small battery, computational and limited range receiving and transmitting capabilities to perform the programmed task. After sensing the data by sensor nodes they send their data to sink. Sensor nodes make a mutual coordination with nearest sensor nodes for data transmission. These sensor nodes behave like a router for transmitting information as shown in **[Fig. 1]**

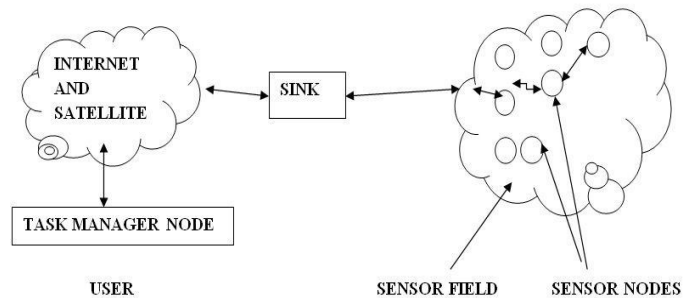

**Fig.1: Wireless Sensor Network**

Power supplied of these sensor nodes is limited. Placements of sensor nodes in improper places and difficulty in changing or recharging batteries in unpractical environment have made researchers to do investigations on reduction of energy consumption. For increasing lifetime, a route selected for data transmission is with low distance from the sink which tends to save the energy consumption and network lifetime increases. Challenging factors during design of routing protocols are like deployment of sensor nodes, network dynamics, energy dissipation, scalability, production cost, network topology, data delivery model, sensor nodes capability, data fusion [19]. Number of routing protocols has been proposed for WSNs. According to Akkaya, K. *et al*. [14] protocols can be classified into three categories.

1. Data Centric routing protocols
2. Hierarchical routing protocols
3. Location based routing protocols
4. Network Flow and QoS-aware Based

Direct Communication between sensor nodes and the base station, as proposed in Direct Transmission (DT) [15] are not encouraged as far sensor nodes from sink die more rapidly. In Minimum Transmission Energy (MTE) [15] sensor nodes near sink die rapidly as these sensor nodes act as relay nodes and they have to transmit large amount of data. These single tier networks can be overloaded with number of transmissions and may cause latency. Main aim of Hierarchical based protocols is to maintain the energy dissipation by using the multi-hop transmissions in clusters. These protocols provide scalability to network which is major design issue for networks. Clustering is proposed for energy efficient protocols which reduce the energy dissipation for long distance transmissions. As shown in **[Fig. 2]** sensor nodes are divided into groups i.e. clusters with a leader (often called cluster head (CH)), elected by the sensors in a cluster or pre-elected by the network designer. Member Sensor nodes communicate information only to cluster heads which process and transfer the aggregated information to Base station. The cluster membership may be fixed or variable. Each clustering algorithm has mainly two phases: Setup phase, election of CH and formation of cluster is performed and Steady state phase data is transmitted from node to CH; CH then aggregates this data and transmits it to BS. Advantages of clustering as compared to other algorithms are:

- Reduces the size of the routing table.
- Reduce the overhead of topology management.
- Reducing the redundant data and reduce latency.
- Decrease system latency and save energy.
- Minimize energy consumption in cluster.

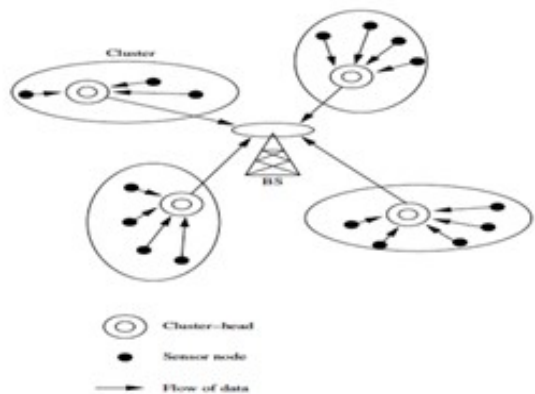

Fig. 2: Clustering in Sensor Network

In Literature, there are two types of networks: Homogeneous Network and Heterogeneous Network. In Heterogeneous networks, sensor nodes are of different type in the sense of size, shape, hardware configuration and the mode of energy supply where in homogeneous networks sensor nodes are identical in these terms. Researchers generally assume that the sensor nodes in wireless sensor networks are homogeneous, but heterogeneous wireless sensor networks are very much useful in real deployments because they are more close to real life situations. There are two types of clustering schemes have been proposed. Firstly, the clustering algorithms applied in homogeneous networks are like the Low-Energy Adaptive Clustering Hierarchy (LEACH)[11], Power-Efficient Gathering in Sensor Information Systems (PEGASIS)[16],Threshold Sensitive Energy Efficient Network(TEEN) [17]. But under the conditions of network heterogeneity this protocol will not be efficient and gives poor performance. Secondly, the clustering algorithms applied in heterogeneous networks are like the Stable Election Protocol (SEP) [12], Energy Efficient Clustering Scheme (EECS) [18], Energy efficient heterogeneous clustered (EEHC) [21], Distributed Energy Efficient Clustering (DEEC) [1].

DEEC [6] is proposed for two-level heterogeneous network where two types of sensor nodes, with different initial energy exist which make network having two levels of heterogeneity. So the resources are having energy heterogeneity. Katiyar *et al*. [13] told that heterogeneity decreases the latency and enhance the network lifetime. DEEC is clustering-based algorithm in which cluster head is selected on the basis of probability of ratio of residual energy of each node and the average energy of the network. The rotating epoch for being cluster heads for each node is different according to its initial energy and residual energy. The node having more energy has more chances to be a cluster head. It prolongs the stability period of the network and more effective messages than other heterogeneous protocols. Many protocols have been derived from DEEC with help of some enhancements and applying advance routing techniques. This paper discuses and compares few hierarchical routing protocols like DEEC, Stochastic-DEEC (SDEEC), Developed-DEEC (DDEEC), Enhanced-DEEC(EDEEC),Stochastic and Balanced-DEEC(SBDEEC), Threshold-DEEC (TDEEC), Enhanced Developed-DEEC(EDDEEC),Hybrid-DEEC (H-DEEC), Balanced Energy Efficient Network Integrated Super Heterogeneous(BEENISH), Hybrid Energy Efficient Reactive (HEER).

The structure of the remaining paper is as follows: Section 2 briefly describes the Distributed Energy Efficient Clustering Protocol. In Section 3, our energy model used in all the variants is described. In Section 4, Performance measures are described which are used in variants. In Section 5, Variants of DEEC are presented. Finally, Section 6 concludes the paper and provides directions for some future work.

## [2] DISTRIBUTED ENERGY EFFICIENT CLUSTERING PROTOCOLS FOR HETEROGENEOUS WIRELESS SENSOR NETWORK

DEEC scheme is energy aware adaptive clustering algorithm which is proposed by Qing *et al.* [1] for two-level and multi-level heterogeneous wireless sensor network. We consider a sensor network with N sensor nodes. For two level heterogeneous network fraction of sensor nodes are considered as advanced sensor nodes (m) with additional energy as compared with normal sensor nodes ($\alpha$). DEEC calculate ideal value of network lifetime which is used to compute the reference energy that each node should expend during a round, to avoid that each node needs to have global knowledge of the network. For cluster head selection in each round 'r', average probability, $P_i$ is calculated which is the ratio of residual energy of $s_i$ node, $E_i(r)$ and average energy of the network, $E_{avg}(r)$. Average energy of network is

$$E_{avg}(r) = \frac{1}{N}\sum_{i=1}^{N} E_i(r)$$

The optimal probability of a sensor node to become a cluster head. $P_{opt}$ can be calculated on the basis of optimum number of clusters $k_{opt}$ as follow:

$$P_{opt} = \frac{k_{opt}}{N}$$

Epoch is number of rounds in which each sensor node become cluster head at least once. DEEC algorithm has rotating epochs for each node according to initial and residual energy of node. The sensor nodes with high energy are more probable to become CH than sensor nodes with low energy. Where sensor nodes in homogeneous network have equal probability to be CH so, $P_i = P_{opt}$. It assure that number of cluster heads formed in each round are $P_{opt} \times N$. But according to DEEC, each node can't have same residual energy after each round. So, this scheme choose different epoch, $L_i$ which is based on residual energy of each node, $E_i(r)$. Average probability of node to be cluster head during epoch can be calculated by $P_i = 1/L_i$ and sensor nodes with high energy have larger value of $P_i$ as compared to the $P_{opt}$ and vice versa. DEEC guarantees optimum number of cluster heads per round per epoch by using average probability for cluster head selection for each node. Net value of cluster heads during each round is equal to $P_{opt} \times N$.

$$\sum_{i=1}^{N} P_i = \sum_{i=1}^{N} P_{opt} \frac{E_i(r)}{E_{avg}(r)} = P_{opt} \sum_{i=1}^{N} \frac{E_i(r)}{E_{avg}(r)} = P_{opt} N$$

Average probability for CH selection in DEEC can be calculated by using $E_{avg}(r)$ and $E_i(r)$

$$P_i = P_{opt}\left[1 - \frac{E_{avg}(r) - E_i(r)}{E_{avg}(r)}\right] = P_{opt} \frac{E_i(r)}{E_{avg}(r)}$$

For cluster head selection, G is the set of sensor nodes eligible to become cluster head at round r. Value of G is reset after each epoch. It considers sensor nodes which have not been cluster head in current epoch. Each eligible node chooses a random number between 0 and 1 during each round. If number is less than threshold, $T(s)$ and sensor node, $s_i$ belongs to set G, it is eligible to become a CH otherwise not.

$$T(s) = \begin{cases} \frac{P_i}{1 - P_i\left(r \bmod \left(\frac{1}{P_i}\right)\right)} & if\ s_i\ \epsilon\ G \\ 0 & otherwise \end{cases}$$

As $P_{opt}$ is reference value of average probability $P_i$. In homogenous networks, all sensor nodes have same initial energy so $P_{opt}$ is used as the reference value for average probability $P_i$. However in heterogeneous networks, the initial energy of sensor node is different with different $P_{opt}$ value. $P_{opt}$ value can be used to calculate the weighted probability of normal sensor nodes and advanced sensor nodes as

$$P_{nrm} = \frac{P_{opt}}{1 + \alpha m},\ for normal nodes$$

$$P_{adv} = \frac{P_{opt}(1 + \alpha)}{1 + \alpha m},\ for advanced nodes$$

Using probability of two types of sensor nodes average probability can be calculated as:

$$P_i = \begin{cases} \frac{P_{opt} E_i(r)}{(1 + \alpha m) E_{avg}(r)} & for normal nodes \\ \frac{(1 + \alpha) P_{opt} E_i(r)}{(1 + \alpha m) E_{avg}(r)} & for advanced nodes \end{cases}$$

This model can be extended for multilevel heterogeneous networks. For multi-level heterogeneous network weighted probability can be changed to

$$P_{multi} = \frac{P_{opt} N(1+\alpha_i)}{N + \sum_{i=1}^{N} \alpha_i}$$

By using above $P_{multi}$ in equation above we can get $P_i$ for multi-level heterogeneous sensor nodes as:

$$P_i = \frac{P_{opt} N(1+\alpha) E_i(r)}{\left(N + \sum_{i=1}^{N} \alpha_i\right) E_{avg}(r)}$$

In DEEC we estimate average energy $E_{avg}(r)$ of the network for any round r. The actual energy of each node will fluctuate around the reference energy $E_{avg}(r)$. Therefore, DEEC guarantees that all the sensor nodes die at almost the same time.

$$E_{avg}(r) = \frac{1}{n} E_{total}\left(1 - \frac{r}{R}\right)$$

R denotes total rounds of network lifetime for finding ideal value of lifetime which can be estimated as follows:

$$R = \frac{E_{total}}{E_{round}}$$

Simulation of DEEC shows that it extends the time from start of simulation to first node died (stability period) as the energy of sensor nodes is used efficiently and more effective messages are transmitted to base station than other heterogeneous protocols. DEEC deals with both two levels and multi-level heterogeneity. DEEC estimates the ideal value of network lifetime which is also used to compute the reference energy that each node expend in each round so it doesn't need any universal knowledge of the network. But there is disadvantage of DEEC that advanced sensor nodes are given more probability to be chosen as cluster head as they have more initial and residual energy so being punished always, particularly when their energy drain and become in range of the normal sensor nodes. That's why advanced sensor nodes die rapidly than normal sensor nodes.

## [3] CHANNEL PROPAGATION MODEL

We describe First order radio model proposed in LEACH [11] and used in DEEC and its variants. In wireless channel, the electromagnetic wave propagation can be modelled as falling off as a power law function of the distance between the transmitter and receiver. In order to achieve Signal-to-Noise Ratio (SNR) in transmitting an L-bit message over a distance d, Energy dissipation by radio model is given as:

$$E_{TX}(l,d) = \begin{cases} LE_{elec} + LE_{fs}d^2 & if\, d \leq do \\ LE_{elec} + LE_{mp}d^4 & if\, d > do \end{cases}$$

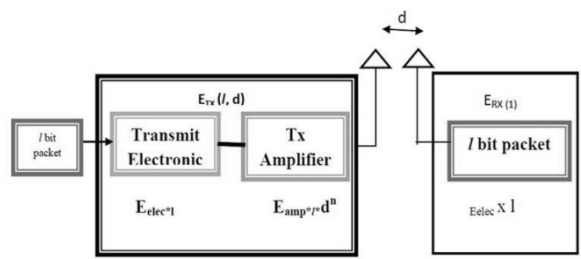

**Fig 3: Radio energy dissipation model**

$E_{elec}$ is energy dissipated per bit to run transmitter ($E_{TX}$) or receiver ($E_{RX}$) circuit. $E_{fs}$ and $E_{mp}$ depend on the radio model used, and d is the distance between the sender and the receiver. Free space model ($d^2$ power loss) which consider direct line-of-sight and two-ray ground propagation is considered if distance between transmitter and receiver. Multi path fading ($d^4$ power loss) channel models is used when distance is more than threshold distance which consider two-ray ground propagation. $d_o$ is threshold distance. By equating the two expressions at $d = d_o$, we have

$$d_o = \sqrt{\frac{Efs}{Emp}}$$

Threshold distance is taken 70m in DEEC. For receiving the data energy dissipation of receiver is $E_{TX} = L.E_{elec}$.

For field A=M×M meter squares and assuming that base station in centre, the energy dissipation by cluster head node during a round can be given as:

$$E_{CH} = L\left[\left(\frac{n}{k}-1\right)E_{elec} + \frac{n}{k}E_{DA} + E_{elec} + E_{fs}d^2_{toBS}\right]$$

First part of equation shows the energy dissipated, $E_{elec}$ by cluster head node to receive $\left(\frac{n}{k}-1\right)$ messages from sensor nodes associated with it. Next is energy dissipated in data aggregation, $E_{DA}$. Then energy used in transmission of data to sink. As sink is in centre it transmits in free space. Equation can be simplified as follow:

$$E_{CH} = L\left[\frac{n}{k}E_{elec} + \frac{n}{k}E_{DA} + E_{fs}d^2_{toBS}\right]$$

Energy dissipation of non-cluster head sensor nodes include energy used in transmitting the data to associated cluster head

$$E_{non-CH} = L[E_{elec} + +E_{fs}d_{toCH}^2]$$

Total energy dissipation in a network is equal to:

$$E_{total} = E_{CH} + \frac{n}{k}E_{non-CH}$$

$$E_{total} = L[2NE_{elec} + NE_{DA} + E_{fs}d_{toCH}^2 + E_{fs}d_{toBS}^2]$$

If the base station is far away the signal need to amplify and total energy dissipation of network will be

$$E_{total} = L[2NE_{elec} + NE_{DA} + E_{fs}d_{toCH}^2 + E_{mp}d_{toBS}^4]$$

As we are assuming that sensor nodes are uniformly distributed, we can calculate

$$d_{toCH} = \frac{M}{\sqrt{2\pi k}}, \quad d_{toBS} = 0.765\frac{M}{2}$$

Optimum number of clusters is very important to calculate. If clusters are not constructed in optimal way, total energy consumption of network increases. Optimum number of clusters,

$$k_{opt} = \frac{\sqrt{N}}{\sqrt{2\pi}}\sqrt{\frac{E_{fs}}{E_{mp}}}\frac{M}{d_{toBS}^2}$$

Energy Model given above is used for all the variants of DEEC.

## [4] PERFORMANCE METRICS

Here we briefly describe all the performance metrics which are used to study and evaluate the performance of different Protocols.

- *Network Lifetime:* It is time interval from start of operation until the death of the last alive node.
- *Stability Period:* It is the time interval from the start of the simulation operation until the death of the first node.
- *Instability Period:* It is time interval from death of first sensor node until the death of last sensor node.
- *Number of alive sensor nodes:* It is a measure that reflects the total number of sensor nodes that has not yet used all of their energy.
- *Network energy consumption:* It is a measure to find the total energy dissipation of the network. It is calculated at each round of the protocol. Less the energy dissipation tends to longer lifetime of network.
- *Throughput:* It is measure the total rate of data sent by sensor nodes of network. Throughput considers data sent from sensor nodes to cluster head and from cluster head to base station.

Longer the lifetime is required in network where the data is required from the network until the death of last node even that data is unreliable. If we need reliable data then stability period should be long and smaller the instability period.

## [5] VARIANTS OF DEEC PROTOCOL:

**[5.1] Developed Distributed Energy Efficient Clustering (DDEEC) Protocol:**
Although the conventional DEEC has many advantages but advanced sensor nodes are always penalized due to high weighted probability of advanced sensor nodes. The problem could be solved by changing the probability of sensor nodes while cluster head selection. So, DDEEC was proposed by Brahim *et al*. [2] which use same method for cluster head selection, average energy of network calculation and estimation of network lifetime.

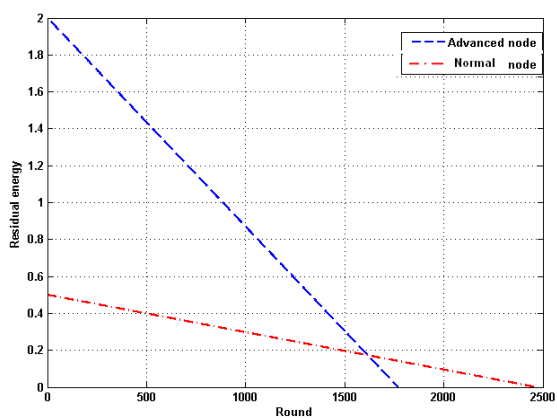

**Fig 4.Variation of residual energy of advanced and normal nodes as shown in [2]**

As in DEEC, we find the sensor nodes with more residual energy are more probable to become cluster head than sensor nodes with less residual energy at round r. In this way, advanced sensor nodes become cluster head more often as compared to normal sensor nodes. A point comes when the residual energy of advanced sensor nodes is same as normal sensor nodes and although decreases than normal sensor nodes as shown in **[Fig. 4]**. Means the advanced sensor nodes continues to be punished and dies more quickly than normal sensor nodes so this is not optimal way for energy distribution. DDEEC avoid this unbalanced distribution by making changes in average probability to save advanced sensor nodes from being punished over and again. DDEEC calculate threshold residual energy, $TH_{\Re}$ which is value of intersection of energies of normal and advanced sensor nodes where $EdisNN$ energy is dissipated by normal node and $EdisAN$ energy is dissipated by advanced node.

$$TH_{\Re} = Eo\left(1 + \frac{aEdisNN}{EdisNN - EdisAN}\right)$$

Let

$$b = \left(1 + \frac{aEdisNN}{EdisNN - EdisAN}\right)$$

Value of first node die is found with variation in function of $b$ and it's found that value of $b$ = 0.7. When energy level of advanced and normal sensor nodes falls down to the limit of threshold residual energy then both type of sensor nodes use same probability to become cluster head. Therefore, CH selection is balanced and more efficient. Average probability $P_i$ for CH selection used in DDEEC can be calculated by using threshold residual energy of network and probability becomes

$$P_i = \begin{cases} \dfrac{P_{opt}E_i(r)}{(1+\alpha m)E_{avg}(r)} & \text{for normal nodes}, E_i(r) > TH_{RE} \\ \dfrac{(1+\alpha)P_{opt}E_i(r)}{(1+\alpha m)E_{avg}(r)} & \text{for advanced nodes}, E_i(r) > TH_{RE} \\ c\dfrac{(1+\alpha)P_{opt}E_i(r)}{(1+\alpha m)E_{avg}(r)} & \text{for advanced, normal nodes}, E_i(r) \leq TH_{RE} \end{cases}$$

$\alpha$ is the amount of energy enhanced of m advanced sensor nodes. *c* is real positive variable which control directly the number of cluster heads. *c* affect the round value of first node dead and value of *c* is taken as 0.02.

So, DDEEC implement a balanced and dynamic probability distribution to spent energy more equitably and to save the energy of advanced sensor nodes which were being punished regularly. Performance of DDEEC is enhanced as compared to DEEC [1] and SEP in terms of stability period that make the network more reliable.

**[5.2] Enhanced Distributed Energy Efficient Clustering (E-DEEC) Scheme:** In real scenario there are more levels of heterogeneity. As when new sensor nodes are deployed some alive sensor nodes are available with different energy. E-DEEC scheme is proposed by Saini *et al*. [3] which consider three level energy heterogeneous networks hence, it increase the level of heterogeneity in network which gives more accurate environment. It contains three types of sensor nodes based on their initial energy: Normal sensor nodes, advanced sensor nodes and Super sensor nodes to improve the energy distribution. E-DEEC improve stability period, number of packets transmitted to base station and number of alive sensor nodes in network as compared to SEP which is also extended to three level heterogeneous network. This scheme shows that DEEC perform well with sensor nodes with more level of heterogeneity. Average probability for cluster head selection, $P_i$ was for two types of sensor nodes which is enhanced to three types of sensor nodes. Value of $P_i$ in E-DEEC is as follows:

$$P_i = \begin{cases} \dfrac{P_{opt}E_i(r)}{(1+m(\alpha+m_o b))E_{avg}(r)} & \text{for normal nodes} \\ \dfrac{(1+\alpha)P_{opt}E_i(r)}{(1+m(\alpha+m_o b))E_{avg}(r)} & \text{for advanced nodes} \\ \dfrac{(1+\alpha)P_{opt}E_i(r)}{(1+m(\alpha+m_o b))E_{avg}(r)} & \text{for super nodes} \end{cases}$$

$m$ is percentage of advanced sensor nodes, $m_o$ is percentage of super sensor nodes from total sensor nodes and $a$ and $b$ are the amounts by which the energy of advanced and super sensor nodes is increased. Simulation results shows that E-DEEC has better performance as compared to SEP in terms of stability period as the energy of network is increased with three level of heterogeneity where the instability period of SEP is longer than E-DEEC as the energy is well distributed and throughput improves with longer lifetime of network. Total energy consumption is also reduced in E-DEEC which helps in increasing the network lifetime.

### [5.3] Enhanced Developed Distributed Energy Efficient Clustering (EDDEEC):

EDDEEC is proposed by Javaid *et al.* [9] which combine the features of E-DEEC and DDEEC proposed earlier. E-DEEC has enhanced the level of heterogeneity but there is still the same problem exist as in two level heterogeneous network. The sensor nodes with more energy having high probability to become cluster head so they are being punished regularly. In E-DEEC, advance and super sensor nodes are chosen more repeatedly as cluster head as they are more probable to be cluster heads so energy of these sensor nodes deplete continuously. To manage this unbalanced condition method proposed in DDEEC is used. Absolute residual energy, $T_{absolute}$ is used to change the probability for cluster head selection. If the energy of sensor nodes decreases as compared to absolute residual energy then the probability becomes equal and till that point probability is weighted. Proposed probabilities for CH selection in EDDEEC are given as follows:

$$P_i = \begin{cases} \dfrac{P_{opt} E_i(r)}{(1 + m(\alpha + m_o b))E(r)} & \text{for normal nodes}, E_i(r) > T_{absolute}, \\ \dfrac{(1 + \alpha) P_{opt} E_i(r)}{(1 + m(\alpha + m_o b))E(r)} & \text{for advance nodes}, E_i(r) > T_{absolute}, \\ \dfrac{(1 + b) P_{opt} E_i(r)}{(1 + m(\alpha + m_o b))E(r)} & \text{for super nodes}, E_i(r) > T_{absolute}, \\ c \dfrac{(1 + \alpha) P_{opt} E_i(r)}{(1 + m(\alpha + m_o b))E(r)} & \text{for normal, advanced, super nodes}, \\ & E_i(r) \leq T_{absolute}, \end{cases}$$

The value of absolute residual energy level, $T_{absolute}$, is written as:

$$T_{absolute} = zEo$$

Where, $z \in (0,1)$. If $z = 0$ then we have traditional E-DEEC. In reality, advanced and super sensor nodes may have not been a CH in rounds *r*, it is also probable that some of them become CH and same is the case with the normal sensor nodes. So, exact value of *z* is not sure. However, through numerous of simulations using random topologies, we try to estimate the closest value of *z* by varying it for best result based on first dead node in the network and find best result for *z* = 0.7. Therefore, $T_{absolute} = (0.7)Eo$

EDDEEC increases the level of heterogeneity which makes it closer to real scenario and distributes the energy consumption of network more efficiently and save the energy of super and advanced sensor nodes by probability update.

### [5.4] Balanced Energy Efficient Network Integrated Super Heterogeneous (BEENISH) Protocol:
BEENISH is proposed by Qureshi *et al.* [10] for efficient energy consumption. As described earlier also in real scenarios wireless sensor networks have various ranges of energy levels. In older schemes two or three energy levels of sensor nodes are considered. Due to random CH selection various levels of energy is created as residual energy of each node can't be same. So, as much more energy levels we quantize and define different probability for every energy level will lead to as much better results and lead to energy efficiency. In proposed scheme sensor nodes are considered with four levels of energies where normal sensor nodes have normal or initial energy, advance nodes have high energy as compare to normal nodes, super have higher energy and ultra-super sensor nodes have highest level of energy. Probability of ultra-super sensor nodes is highest so energy consumption is equally distributed. The probabilities for four types of sensor nodes are as follows:

$$P_i = \begin{cases} \dfrac{P_{opt} E_i(r)}{\left(1 + m\left(\alpha + m_o(-\alpha + b + m_1(-b + u))\right)\right) E(r)} & \text{for normal nodes} \\ \dfrac{(1+\alpha) P_{opt} E_i(r)}{\left(1 + m\left(\alpha + m_o(-\alpha + b + m_1(-b + u))\right)\right) E(r)} & \text{for advanced nodes} \\ \dfrac{(1+b) P_{opt} E_i(r)}{(1 + m(\alpha + m_o(-\alpha + b + m_1(-b + u)))) E(r)} & \text{for super nodes} \\ \dfrac{(1+\mu) P_{opt} E_i(r)}{\left(1 + m\left(\alpha + m_o(-\alpha + b + m_1(-b + u))\right)\right) E(r)} & \text{for ultra}-\text{super nodes} \end{cases}$$

BEENISH increases the network lifetime by increasing the level of heterogeneity as the sensor nodes with more energy exist in network as ultra-super sensor nodes. These sensor nodes provide more reliability in the network and increase the network lifetime as compared to DEEC, E-DEEC and DDEEC. With the more networks lifetime number of alive sensor nodes are also more that transmits data to base station continuously.

**[5.5] Threshold Distributed Energy Efficient Clustering (TDEEC) Protocol:**
Although DEEC is energy efficient protocol which use the energy of sensor nodes efficiently, as sensor nodes are given duty according to their residual energy. Main aim of TDEEC is to use energy of sensor nodes more efficiently by changing threshold. Saini *et al*. [4] proposed TDEEC, threshold value is adjusted by introducing residual energy of a node and average energy of network of that round with respect to optimum number of cluster heads. Based upon that threshold value; a node decides whether to become a cluster head or not. Other cluster head selection procedure is same as earlier. Threshold value proposed by TDEEC is given as follows:

$$T(s) = \begin{cases} \dfrac{P}{1 - P\left(r \bmod\left(\dfrac{1}{P}\right)\right)} * \dfrac{Residual\ Energy\ of\ a\ node * K opt}{Average\ Energy\ of\ network} & \\ & if\ s \in G \\ 0 & otherwise \end{cases}$$

Enhanced threshold is implemented for various levels of heterogeneity and with all types of heterogeneities TDEEC is efficient as compared to SEP and DEEC as the selection of sensor nodes become more efficient. Simulation shows that sensor nodes remain alive for long in TDEEC and number of packet transmitted are increased and stability period also improve.

**[5.6] Stochastic Distributed Energy Efficient Clustering (SDEEC):** SDEEC is a self-organized network with dynamic clustering concept. Elbhiri *et al.* [5] Introduces a dynamic method where the probability for cluster head selection is more efficient. The key idea behind SDEEC is to reduce the number of intra-clusters transmissions. This scheme can be used when the objective is to collect the maximum or minimum data values in a field. Network lifetime is improved because of reduction in number of transmissions which leads to saving the energy of the sensor nodes.

According to this protocol, Cluster head always keep it's receiver on for receiving the data from the associated sensor nodes in the cluster. Cluster head broadcast its sensed information and only sensor nodes with the significant data transmit there data to cluster head and other sensor nodes must be in sleep mode. So total number of reception and transmissions are reduced which save the energy consumption. Then cluster head do fusion/signal processing of information and send fused information to base station. Let suppose that each node $i$ on the network has a probability $S_i$ to have the searched value in the considered cluster. The total energy consumed $E_{total}$ in the network to transmit the significant data to the sink is:

$$E_{total} = K_{opt}\left(E_{toBS} + \sum S_i E_{toCH}\right)$$

Where,

$E_{toBS}$ is the energy consumed when the cluster head transmit data to the base station.

$E_{toCH}$ is the energy consumed to transmit data from node *i* to the cluster head.

$K_{opt}$ is optimal number of cluster heads.

Equation below shows total energy consumed by round is largely reduced as compared to DEEC protocol

$$E_{total} = LK_{opt}\left[E_{mp}d_{toBS}^4 + E_{elec} + \left(E_{elec} + E_{DA} + E_{fs}d_{toCH}^2\right)\sum S_i\right]$$

Simulation results show that SDEEC performs better than SEP and DEEC in terms of stability period. As the sensor nodes remain alive for more number of rounds so data transmission is more. The number of data transmissions and receptions are decreased with significant data collection hence; total energy consumption is also decreases.

**[5.7] Stochastic and Balanced Distributed Energy-Efficient Clustering (SBDEEC):** Elbhiri *et al.* [6] Proposes a balanced and dynamic method where the cluster head election probability is more efficient. It use stochastic scheme proposed in SDEEC to reduce the number of data transmissions and receptions to enhance the network lifetime and balance the energy of network by using the threshold residual energy to make the weighted probability more efficient as proposed by DDEEC protocol. With Balanced energy stability period increases as compared to SEP and DEEC but the network lifetime was decreased. When the network is both the balanced and stochastic the network lifetime increases and stability period is also improved. This protocol saves the energy of network where the energy of SEP and DEEC drops rapidly.

**[5.8] Hybrid-Distributed Energy Efficient Clustering: Towards Efficient Energy Utilization in wireless sensor network:** H-DEEC is proposed by Khan *et al*. [7] for two level heterogeneous networks. Each node in the network is aware about all the other sensor nodes energy level and position. The H-DEEC scheme is combination of two scenarios;

1. Clustering
2. Chain construction.

Clustering scheme follow the thoughts of DEEC. Estimation of average energy of network, probability of sensor nodes to become cluster head is based on residual energy, cluster head selection algorithm is same as in DEEC. Cluster heads aggregate the data and send that data to nearest high energy node (called beta node). PEGASIS [16] is chain based routing scheme where the sensor nodes send data to their nearest sensor nodes and a leader node send data to base station. Similarly, Beta sensor nodes collect data from the cluster heads and send to other beta node. Chain formation is initiated by base station which marks the farthest node and this node finds its neighbour and so on. Beta sensor nodes are connected by greedy algorithm. Beta node with least distance from base station is chosen as a leader node. Data transmission is done by multi-hopping towards the base station. This improves delivery of data as cluster heads don't transmit directly and saves a significant amount of energy as compared with the other routing protocols. This approach will distribute load evenly among the sensor nodes, due to which instability time decreases. To overcome the deficiencies of H-DEEC Multi-edged Hybrid–DEEC (MH-DEEC) is proposed. This differs in chain formation where the clustering is same as H-DEEC. Unlike H-DEEC, chain construction process is done as in improved energy efficient PEGASIS-based (IEEPB)[20] protocol where initiation is done by BS by marking the farthest beta node from BS and every node connects itself to its nearest neighbour node, doesn't matter whether that node is already connected or not which leads to formation of a Multi-Edged Chain. This solves the problem of long link. Chain leader is chosen using weighted method on the basis of residual energy and distance of each beta node from base station. Beta node with minimum weight is chosen as leader for that round. Weight is calculated with $w_1$ and $w_2$ weight factors:

$$W_i = w_1 E_p + w_2 D_{toBS}$$

Energy parameter $E_p$ is calculated as follows:

$$E_p = \frac{E_{initb}}{E_{i-b}(r)}$$

Where $E_{initb}$ represent initial energy of the beta node and $E_{i-b}(r)$ is the residual energy.

Distance parameter $D_{toBS}$ is calculated by multi-path model given as:

$$D_{toBS} = \frac{d_{toBS}^4}{d_{avg}^4}$$

Where $d_{toBS}$ is distance of beta node from BS and $d_{avg}$ represents the average distance between beta sensor nodes and BS. Finally, the comparison in simulation results with other heterogeneous protocols show that, MH-DEEC and H-DEEC achieves longer stability time and network life time due to efficient energy utilization. Again as in DEEC, beta sensor nodes are always burdened.

**[5.9] Hybrid Energy Efficient Reactive (HEER) Protocol:** In reactive protocols, sensor nodes react immediately to sudden and drastic changes in the value of a sensed attribute. As such, they are well suited for time critical applications. Other protocols proposed earlier are proactive protocols which are well suited for the applications requiring periodic data monitoring. HEER is a reactive protocol proposed by Javaid *et al*. [8] which improves the stable region for clustering hierarchy process for a homogeneous and heterogeneous environment. Clustering in proposed scheme is same as in DEEC Every node select itself as CH on the basis of its Initial and residual energy similar to that of DEEC. It does not require any global knowledge of energy at any election round. Data transmissions occur as follows:

1. When cluster formation is done, the CH transmits two threshold values, i.e. HT (Hard threshold) and ST (Soft threshold). The sensor nodes sense their environment repeatedly and if a parameter from the attributes set reaches its HT value, the node switches on its transmitter and transmits data to the CH. The CH aggregates and transmits data to base station.
2. The Current Value (CV), on which first transmission occurs, is stored in an internal variable in the node called Sensed Value (SV). The node, then again starts sensing its environment until the CV differs from SV by an amount equal to or greater than ST , $CV - SV \geq ST$ . When this condition becomes true, the node again switches on its transmitter and sends data to CH. It further reduces the number of transmissions. The CH then transmits data to base station.

HEER performs best for time critical applications in both homogeneous and heterogeneous environment. Message transmission consumes much more energy than data sensing. So, even though the sensor nodes sense continuously, the energy consumption in this scheme can potentially be much less than in the proactive network, because data transmission is done less frequently. It increases the stability period and network lifetime. The main drawback of this scheme is that, if the thresholds are not reached, the sensor nodes will never communicate; the user will not get any data from the network at all and will not come to know even if all the sensor nodes die. Thus, this scheme is not well suited for applications where the user needs to get data on a regular basis. Another possible problem with this scheme is that a practical implementation would have to ensure that there are no collisions in the cluster. Time division multiple access (TDMA) scheduling of the sensor nodes can be used to avoid this problem. This will however introduce a delay in the reporting of the time-critical data. Code division multiple access (CDMA) is another possible solution to this problem.

## [6] CONCLUSION

Presence of heterogeneity improves the reliability and lifetime of wireless sensor network. Heterogeneous wireless sensor networks are more suitable for real life applications. Energy optimization is challenging issue in WSNs. Clustering is effective way to use energy efficiently. In this paper we surveyed Distributed energy efficient clustering algorithm and its variants. DEEC is energy aware adaptive technique to use the energy of sensor nodes efficiently. Variants of DEEC show new concepts and techniques which improve reliability, lifetime and throughput. Variants work with different levels of heterogeneity. Although these variants show improvements certainly further energy improvement is possible in future by researchers. Improvement is also possible in many aspects like sensor nodes electronics, sensor nodes deployment management, effective and energy efficient routing protocols selection for WSNs according to requirements of application. There is possibility to combine Energy efficient protocol with Quality of Service (QoS) protocols which tends to reduce the latency in network.